\documentclass{kluwer}    % Specifies the document style.
\usepackage{graphicx}
\usepackage[a4paper]{hyperref}
\newdisplay{guess}{Conjecture}

\begin{document}                                                                                   
\begin{article}
\begin{opening}         
\title{Elliptical galaxy nuclei activity powered by infalling globular 
clusters
%\thanks{...}
}
\author{Roberto \surname{Capuzzo-Dolcetta}}
\runningauthor{Roberto Capuzzo-Dolcetta}
\runningtitle{Elliptical galaxy nuclei activity}
\institute{Dep. of Physics, Univ. of Roma, La Sapienza}
%\date{...}

\begin{abstract}
Globular cluster systems evolve, in galaxies, due to internal and external dynamics and tidal
phenomena. One of the causes of evolution, dynamical friction, is responsible for the
orbital decay of massive  clusters into the innermost galactic regions.
It is found that these clusters are effective source of matter to feed a central
 galactic black hole such to make it grow and shine as an AGN.

\end{abstract}

\end{opening}           

\section{Globular cluster systems in galaxies}
                    % Produces section heading.  Lower-level
                    % sections are begun with similar 
                    % \subsection and \subsubsection commands.
Thanks to the high resolution of HST, as well of the power of large ground based telescopes,
good data are available about the characteristics of globular clusters (GCs) in galaxies 
(see, e.g., Forbes et al. 1996; Elson et al 1998; Grillmair et al. 1999).
\par \noindent One general conclusion is that the Globular Cluster Systems (GCSs) are, usually, less centrally concentrated than
the underlying star distribution (Lauer \& Kormendy 1986,  Harris 1986,
Harris et al. 1991, Capuzzo-Dolcetta \& Vignola 1997, Capuzzo-Dolcetta
\& Tesseri 1999, Capuzzo-Dolcetta \& Donnarumma 2001). The simplest explanation is that globular 
clusters and halo-bulge stars, being coeval, formed with the same radial distribution in the galaxy
and the present different distribution  has been reached by the following evolution of the 
GC component. This evolution is caused by
the various mechanisms acting on GCs as galaxy satellites (mainly dynamical friction and
tidal interaction with the overall halo-bulge-disk stellar and gaseous component).
Many authors have studied the problem of evaluating the role of these evolutionary
phenomena by different points of view (Fall \& Rees 1977;
Capuzzo-Dolcetta \& Tesseri, 1997, 1999; Gnedin \& Ostriker 1997;
Murali \& Weinberg 1997a, 1997b; Baumgardt 1998;
Vesperini  2001).
In particular, Pesce, Capuzzo-Dolcetta \& Vietri (1992) have shown how the role of
dynamical friction
is of enormous importance in the evolution of GCs in triaxial galaxies. Capuzzo-Dolcetta (1993)
 developed a model
to study the evolution of GCSs subjected to both dynamical friction and tidal interaction with
a massive galactic nucleus and found that sufficiently massive GCs in a triaxial galaxy,
self-consistently modeled in the way   described by Schwarzschild (1979) and Merritt (1980)
 lose enough orbital energy ($E$)
and angular momentum ($L$) to be confined in a short time in the innermost region of the galaxy. 
These clusters form
a sort of ~\lq supercluster\rq~ which is a large mass {\it reservoire}
for a compact object (a primordial black hole?) sitting there. Whenever the mass density of the
supercluster rises over a certain threshold, the matter accretion onto the black hole (b.h.) may be
the cause of its growing up to super-massive black hole size and of its activity as
an AGN. Of course, when the mass of the b.h. gets
sufficiently large it starts to shatter not only the looser, usually lighter, GCs but also the
few remaining infalling massive clusters, so to halt the b.h. mass growth to an almost steady value.
This means that the mechanism of b.h. mass growth is a {\it self-regulating}.
These general results have been discussed in various papers 
(see for instance Capuzzo-Dolcetta \& Miocchi 1998, Capuzzo-Dolcetta 2002) starting from
Capuzzo-Dolcetta (1993). :
\par
Consequently, it appears clearly defined this {\it scenario}
for the   evolution of the GCS in a triaxial galaxy:
\par (i) massive GCs on {\it box} orbits (in triaxial galaxies) or, equivalently, on low $L$ orbits in
 axisymmetric galaxies  lose their orbital energy rather quickly;
\par (ii) after $\sim 500$ Myr many GCs are limited to move in the inner galactic
region where they can merge and form a {\it supercluster};
\par (iii) stars of the supercluster buzz around the nucleus where they may be
captured by a b.h. seed there;
\par (iv) consequently, energy is extracted from the gravitational field: part of it goes into e.m.
radiation inducing an AGN activity while part increases the b.h. mass.

\section{Black hole growth and nuclear activity}
Consider three well studied galaxies having a firm evidence of presence
of a nuclear b.h.: our Milky Way, M 31 and the giant elliptical M 87 in Virgo.
The modern estimates of the central black hole masses  available in the literature
are, respectively, $2.6\times 10^6$
M$_\odot$, $6.2 \times 10^7$ M$_\odot$ and $3.6 \times 10^9$ M$_\odot$.
In the frame of the presence of a dense stellar system around the black hole, the
simple process of spherical accretion of stars buzzing there around determines a threshold of
 mass density ($\rho_c$) of the nuclear cluster to allow an emission as an AGN, once that the minimum
mass accretion rate of 1 M$_\odot$ yr$^{-1}$ to sustain an AGN activity
is assumed.
\par\noindent The expression of the spherical mass accretion rate is (Capuzzo-Dolcetta 2002):

$$
\dot m (M_\odot yr^{-1})=
8.2\times 10^{-16} m_{bh}^{4/3}m_*^{-1/3} \rho_* R_*<v_*>^{-1}, 
~~r_d=r_t,$$
or 
$$\dot m (M_\odot yr^{-1})=2.8\times 10^{-21} m_{bh}^2 \rho_* <v_*>^{-1}, ~~
r_d=r_S,
$$

where: the index $*$ refers to values of the supecluster stars; $r_t$ and $r_S$ are the ~\lq tidal\rq~ and ~\lq Schwarzschild's\rq~ radius, respectively;
$r_d=max\{r_t,r_S\}$ is the ~\lq destruction radius\rq~; the quantities $m_*$, $m_{bh}$, and 
$R_*$ are in solar units, $\rho_*$ in M$_\odot$ pc$^{-3}$;
and $<v_*>$ (the internal stellar mean velocity) in km s$^{-1}$. Assuming $<v_*>$=10 kms$^{-1}$ for
we get $\rho_c= 3.4\times 10^7$ M$_\odot$ pc$^{-3}$ for the MW, $\rho_c=5\times 10^5$
M$_\odot$ pc$^{-3}$ for M 31 and $\rho_c=300$ M$_\odot$ for M 87.
The value of $\rho_c$ for M87 is clearly a normal-low value for
the central density of galactic GCs, while the $\rho_c$ of M 31, even if large, is compatible with the estimated central values
of some non core-collapsed GCs (see the Harris web Catalog of parameters for Milky Way globular clusters, 
feb. 2003, http://physun.physics.mcmaster.ca/\~harris/mwgc.dat).
The value of $\rho_c$ of the MW is high if compared with typical, non-collapsing
globular clusters but totally compatible with the merging scenario previously sketched.
\noindent Actually, a density of $3.4 \times 10^7$ M$_\odot$ pc$^{-3}$ may be reached by a pure (linear) merging
of abput one hundred GCs typical in their mass and linear sizes
(total mass of few $10^5$ M$_\odot$ and half mass radius of a couple of pc). 
\begin{figure*}
   \centering
   \resizebox{\hsize}{!}{\rotatebox[]{0}{\includegraphics{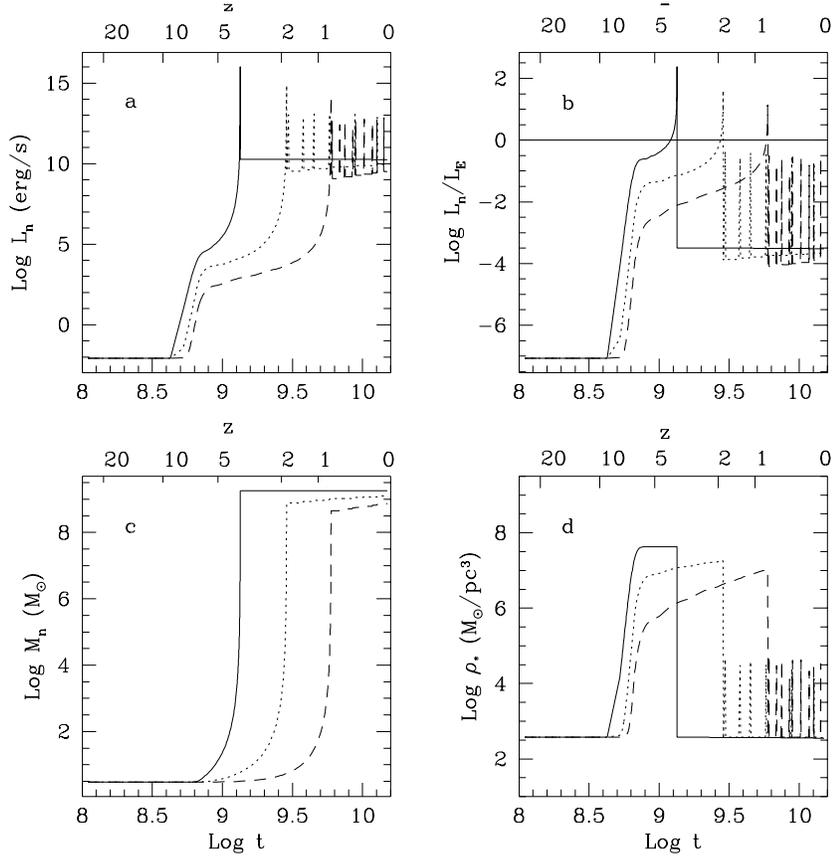}}}
   \caption{This figure refers to the model with M$_{n0}=$3 M$_\odot$. Panel a: time evolution of the galactic nuclear
luminosity induced by GC mass accretion;
solid line refers to the {\it cold} GCS; dotted to the {\it normal} GCS; dashed to the {\it hot} GCS
(time is in yr in Log scale; the upper abscissa is the corresponding red-shift in
the Einstein-De Sitter model). Panel b: nuclear luminosity in units of Eddington's luminosity. Panel c:
time evolution of the galactic nuclear (black hole) mass. Panel d: time evolution of the supercluster
star mass density. }
              \label{lum0}
    \end{figure*}

  \begin{figure*}
   \centering
   \resizebox{\hsize}{!}{\rotatebox[]{0}{\includegraphics{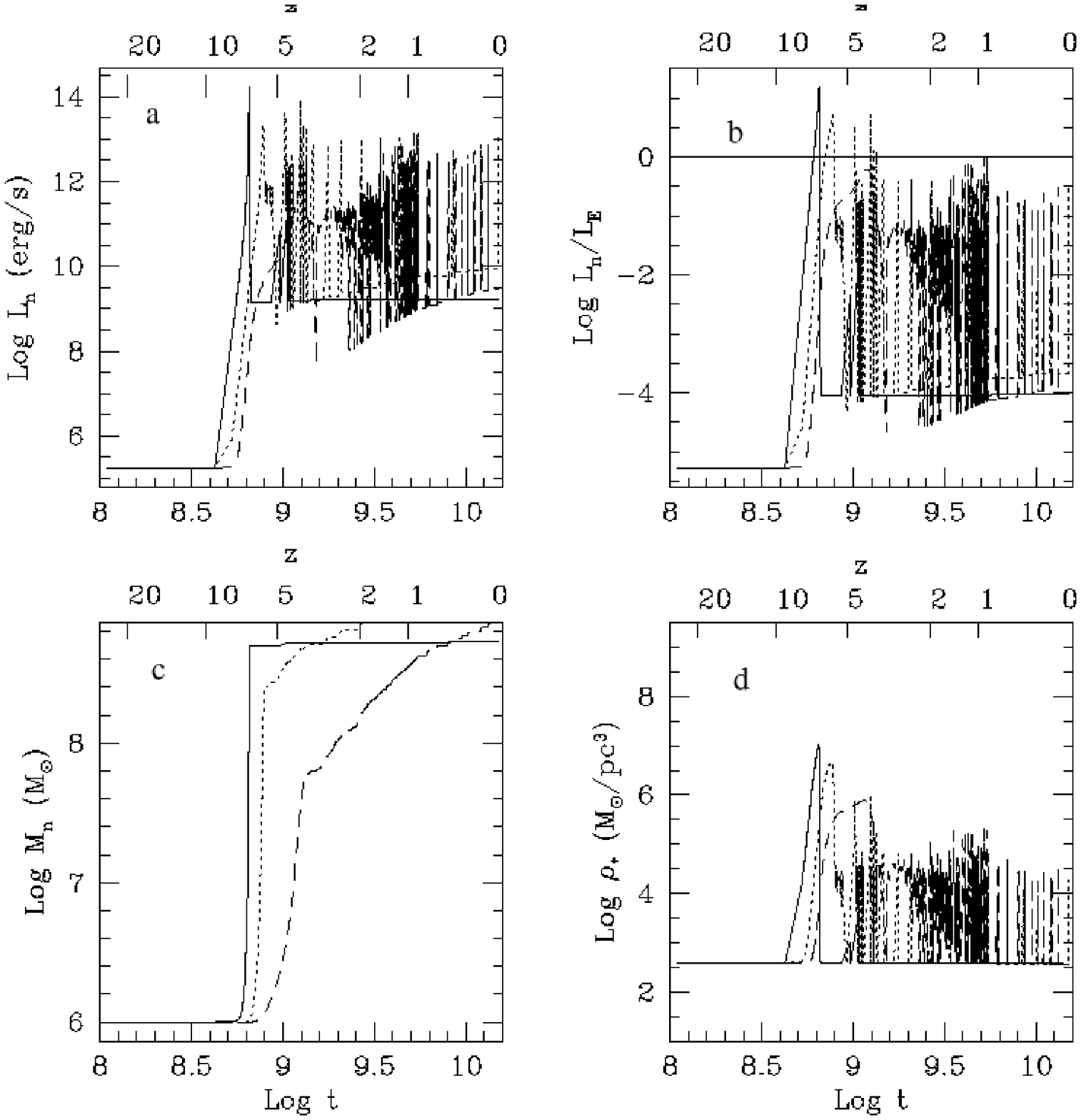}}}
   \caption{As in Fig. 1, for the model with M$_{n0}=$10$^6$ M$_\odot$.}
              \label{lum6}
    \end{figure*}
%\begin{figure*}
%   \centering
%    \includegraphics[width=7 cm]{figure/fig1.eps}
   %\resizebox{\hsize}{4truecm}{\rotatebox[]{0}{\includegraphics{figure/fig1.eps}}}
   %\includegraphics{empty.eps}
   %\includegraphics{empty.eps}
%   \caption{ Bi-logarithmic plot of the curve (solid line) $\rho_*<v_*>^{-1}$
%(in units M$_\odot$ pc$^{-3}$ (kms$^{-1}$)) that corresponds to
%an accretion rate $\dot m = 1$ M$_\odot$ yr$^{-1}$
%in function of the black hole mass $m_{bh}$.
%The dashed line is the ~\lq collisional \rq~ curve (see
%Capuzzo-Dolcetta 2002). }
%              \label{accretion}
%    \end{figure*}
\par Given this plausibility of the accretion-emission mechanism, we report here of some
results that will be presented in a more detailed way in a forthcoming paper.
To employ the  model of evolution of a GCS developed by
Capuzzo-Dolcetta (2001, 2004), we refer to the triaxial galactic model developed by Schwarzschild (1979) and
Merritt (1980),
The system is composed by $N_0=1000$ single mass ($M_0=2\times 10^6$ M$_\odot$) GCs whose
orbital velocity dispersion ($\sigma$) defines a ~\lq normal\rq~ system, as that whose
$\sigma =330$ km s$^{-1}$ is equal to the average $\sigma$ of stars generating the
galactic potential, while the choice of
$\sigma =165$ km s$^{-1}$  corresponds to a dynamically ~\lq cold\rq~ system and
$\sigma =660$ km s$^{-1}$ to a ~\lq hot \rq~ system.
Two different initial masses for the nuclear  galactic b.h. are considered: $M_{n0}= 3$ M$_\odot$ 
and   $M_{n0}= 10^6$ M$_\odot$.
 
Fig. 1 shows the time behaviour of the nucleus luminosity and mass (L$_n$ and M$_n$), as well as
of the super-cluster stellar density by mass ($\rho_*$). The density grows
rapidly up to a plateau at $\rho_* \sim 4.3 \times 10^7$ M$_\odot$ pc$^{-3}$
in the case of the colder GCS; in the other two cases studied, $\rho_*$ shows a similar (later) quick growth
up to a maximum ($\sim 2 \times 10^7$ M$_\odot$ pc$^{-3}$ and
$\sim 1.2 \times 10^7$ M$_\odot$ pc$^{-3}$ for $\sigma = 330$ km s$^{-1}$ and
$\sigma = 660$ km s$^{-1}$, respectively) followed
by a slower increase regime whose slope is greater for
greater $\sigma$'s. These behaviours of $\rho_*$ have a
correspondence in the time behaviour of L$_n$ (Fig. 1 a,b) that shows a
fast growth followed by a more gentle rise up to a short-duration super-Eddington
peak followed by a sudden fall to a level determined by the accretion of the
black hole by  stars of the bulge. The interval of redshift of the luminosity peak is
$0.75\leq z_{peak} \leq 3.7$, in the assumption of an Einstein-de Sitter universe.
 Notice (Fig. 1 c) that in  the interval of time from $1.3$ Gyr ({\it cold} GCS) to $6$ Gyr  ({\it hot} GCS) the black hole mass grows from the stellar size  up to the values of very massive black holes:
$1.8 \times 10^9$ M$_\odot$, $7.5 \times 10^8$  M$_\odot$ and $7.4 \times 10^8$
M$_\odot$, in the three cases studied (cold, normal, hot). Note that a factor 4 in 
$\sigma$ corresponds to just a factor 2 in the final M$_n$.
\par\noindent Fig. 2 refers to the results of the model with $10^6$ M$_\odot$ as initial value of
the central black hole mass. As expected, the AGN activity is shifted backward in time
($6.6\times 10^8$ yr $\leq t_{peak} \leq 1.3$ Gyr, corresponding to
$7.6 \leq z_{peak} \leq 3.5$).  The maximum luminosity reached by the AGN is lower than 
in the case of a b.h. seed of stellar origin; this because the massive b.h. is able to swallow stars of the
forming supercluster before it has grown up to very large densities.
Actually, as seen in Fig. 2 d, the $\rho_*$ profile does not show the regular two-slope
behaviour of the case of an initial small black hole.
In the case of $M_{n0}=10^6$ M$_\odot$, the time evolution of the nucleus mass is such that the colder GCS induces, as usual,
the fastest growth while the value of the black hole mass reached at the, almost, steady state
is smaller than in the case of greater average orbital energy of GCs (see Fig. 2 c).
%\begin{figure*}
%   \centering
%   \resizebox{\hsize}{!}{\includegraphics[clip=true]{figure/fig3a.eps}
%   \includegraphics[clip=true]{figure/fig3b.eps}}
%   \caption{Time evolution of the ratio of the number of ~\lq surviving\rq~ clusters  to the initial
%total number for the cases of M$_{n0}=3$ M$_\odot$ (panel a) and M$_{n0}=10^6$ M$_\odot$ (panel b).
%Panels c and d refer to the ratio of the number of clusters lost due to dynamical friction and
%nucleus tidal disruption, in the two cases as above, respectively. Lines as in Figs. 2 and 3.}
%              \label{lum6}
%    \end{figure*}
The explanation for this is that in the case of the cold GCS the rapid fall of GCs in the central region causes
the subsequent quick  growth of both M$_n$ and L$_n$:  $M_n$ reaches rapidly a value large enough to shatter
tidally
the incoming, lighter GCs. This means that there is no more way to accrete furtherly the b.h.
after the fast consumption of the residual supercluster.
 In this case, the galaxy is spoiled of all its GCs in a time much shorter than the
Hubble time. Among the cases studied here, only the cold and normal GCSs in the case of
M$_{n0}=10^6$ M$_\odot$  suffer of the nucleus tidal erosion, able to destroy $74\%$ and $3 \%$ of the
initial population, respectively.
An interesting result is that the values of the steady (final)  black hole mass are
quite  similar in the various cases studied here: the range of values of the present day
central black hole is $0.7 \div 2 \times 10^9$ M$_\odot$.

\section{Conclusions}
The difference observed among the radial distributions of the family of globular clusters and
bulge-halo stars in galaxies  can be explained in terms of evolution of the globular cluster system.
The evolution of massive clusters is dominated by dynamical friction,
whose effect is enhanced in triaxial galaxies, while tidal effects are relevant
just when a very massive nucleus is initially present or evolutively accreted.
The dynamical friction braking is able to carry a huge amount of mass in the inner galactic
regions in less than 1 Gyr: this provides mass and gravitational energy to a central black hole to 
grow to and radiate  as AGN. 
\par Of course, all these results should be confirmed by direct simulations of globular cluster merging
in the inner galactic regions (work started by Capuzzo-Dolcetta \& collaborators via direct
N-body simulations and presently under way) as well as by a larger set of self-consistent triaxial galactic models,
without (Capuzzo-Dolcetta \& Vicari 2004) and with a density cusp (Capuzzo-Dolcetta, Merritt \& Vicari
2004). The preliminary results seem to confirm the general validity of the results presented here.

\end{article}
\end{document}